\documentclass{PoS}

\usepackage{graphics,psfrag}
\usepackage{amsmath,epsf,epsfig,float,amssymb,latexsym}

\newcommand{\val}{\vec{\alpha}}

\newcommand{\vp}{{\mathbf{p}}}
\newcommand{\vq}{{\mathbf{q}}}

\newcommand{\bg}{\begin{align}}
\newcommand{\eeg}{\end{align}}
\newcommand{\be}{\begin{equation}}
\newcommand{\ee}{\end{equation}}
\newcommand{\ba}{\begin{eqnarray}}
\newcommand{\ea}{\end{eqnarray}}

\newcommand{\nn}{\nonumber}

\newcommand{\vs}{\vspace{-0.2cm}}

\newcommand{\ep}{\epsilon}

\title{Chiral effective Field Theory for Nuclear Matter}

\ShortTitle{Chiral EFT for nuclear matter}

\author{\speaker{J.~A.~Oller}\thanks{I would like to thank my collaborators A.~Dobado, A.~Lacour, F.~Llanes and U.-G.~Mei{\ss}ner for their collaboration in several parts of the results presented.}\\
        Departamento de F\'{\i}sica, Universidad de Murcia, E-30071 Murcia, Spainn\\
        E-mail: \email{oller@um.es}}

\abstract{We review on a  chiral power counting scheme for in-medium chiral
perturbation theory with nucleons and pions as degrees of freedom \cite{ref}.  It allows for a systematic expansion taking into 
account local as well as pion-mediated inter-nucleon interactions. Based on 
this power counting, one can identify classes of non-perturbative diagrams that require
a resummation. 
We then calculate the nuclear matter energy density for the symmetric  and purely neutron matter cases up-to-and-including next-to-leading order (NLO), in good agreement with sophisticated many-body calculations. Next, the neutron matter equation of state is applied to calculate the upper limit for neutron stars, with an upper bound around 2.3 solar masses,  large enough to accommodate the most massive neutron star observed until now. We also apply our equation state to constraint $G_N$ in exceptionally large gravitational fields. }

\FullConference{Sixth International Conference on Quarks and Nuclear Physics,\\
		April 16-20, 2012\\
		Ecole Polytechnique, Palaiseau, Paris}

\begin{document}

\section{Introduction}
\def\theequation{\arabic{section}.\arabic{equation}}
\setcounter{equation}{0}
\label{sec:int}

 An interesting achievement in nuclear physics would be the calculation of atomic nuclei 
and nuclear matter properties from microscopic inter-nucleon forces in a
systematic and controlled way. This is a
non-perturbative problem involving the strong interactions.
 In the last decades, Effective Field Theory (EFT) has proven to be
an indispensable tool to accomplish such aim. 
In this work  we employ  Chiral Perturbation Theory (CHPT) to nuclear 
systems \cite{wein,wein1,wein2}, with nucleons and pions as the pertinent degrees of freedom.
For the lightest nuclear systems with two,
three and four nucleons, it has been successfully applied 
\cite{Epelbaum:2008ga}. 
 For heavier nuclei one common procedure is to employ the chiral
 nucleon-nucleon  potential with standard  many-body 
methods, sometimes supplied with renormalization group techniques~\cite{shaefer}, or in lattice calculations \cite{lee}.
 We have recently derived \cite{ref} a chiral power counting  in nuclear matter that takes into account local multi-nucleon interactions simultaneously to  pion-nucleon interactions. 
Many present applications of CHPT to nuclei and nuclear matter only consider
meson-baryon chiral Lagrangians (see e.g. \cite{Epelbaum:2008ga} for a summary),
without constraints from free nucleon-nucleon scattering.
Our novel power counting was applied in ref.\cite{anals}, among other problems, to 
determine the nuclear matter energy per baryon. We elaborate on these results here. Next, 
we apply them to pure neutron matter case and study the upper limit of a  
neutron star mass. We see that our results can accommodate the recently observed neutron star with a mass $(1.97\pm 0.04)~M_\odot$ \cite{demoret}, the largest one confirmed until now. 
We also use our equation of state to constraint the running of the gravitational constant $G_N$ with the gravitational field intensity, exceptionally large
 (around $2\times 10^{12}~\rm{m/s}^2$) inside a neutron star of  2 solar masses. In this case we  show that $G_N$ cannot exceed its value on  Earth by more than a 12$\%$ \cite{felipe}.

\section{Chiral Power Counting}
\def\theequation{\arabic{section}.\arabic{equation}}
\setcounter{equation}{0}
\label{sec:pw}

Ref.\cite{prcoller}  establishes the concept of an ``in-medium generalized vertex'' (IGV).  
Such type of  vertices result because
one can connect several bilinear vacuum vertices through the exchange of baryon
propagators with the flow through the loop of one unit of baryon number,
contributed by the nucleon Fermi seas.  At least one is needed because otherwise we would have a vacuum closed
nucleon loop that in a low energy EFT  is buried in the 
chiral higher order counterterms. 
 It was also stressed in ref.\cite{annp} that within a nuclear environment a nucleon
 propagator could have a ``standard'' or ``non-standard'' chiral counting. To see
 this note that a soft momentum $Q \sim p$, related to pions or external sources
 can be associated to any of the
 vertices.
  Denoting by $k$ the on-shell four-momenta associated with one
 Fermi sea insertion in the IGV, the four-momentum
 running through the $j^{th}$ nucleon propagator can be written as $p_j=k+Q_j$.
 If $Q_j^0={\cal O}(m_\pi)={\cal O}(p)$  one has the standard counting so that the baryon propagator scales as ${\cal O}(p^{-1})$. However,  
if $Q_j^0$ is  of the order of a kinetic nucleon energy in the nuclear medium 
then  the nucleon propagator should be counted as ${\cal O}(p^{-2})$.
This is referred as the ``non-standard'' 
 case \cite{annp}.  In order to treat  chiral Lagrangians
with an arbitrary number of baryon fields (bilinear, quartic, etc) ref.\cite{ref} considered firstly
bilinear vertices like in refs.\cite{prcoller,annp},  but now  the additional 
exchanges of  heavy meson fields of
any type are allowed. The latter  should be considered as
merely auxiliary fields that allow one to find a tractable
representation of the multi-nucleon interactions that result when  the
masses of the heavy mesons tend to infinity.
 These heavy meson fields are 
denoted in the following by $H$, 
and a heavy meson propagator is counted as ${\cal O}(p^0)$ due to their large masses. On the other hand, ref.\cite{ref} takes  the non-standard counting case from the start
and  any nucleon propagator is considered as  ${\cal O}(p^{-2})$.
In this way, no diagram whose chiral order is actually lower
than expected if the nucleon propagators were counted assuming the standard rules is lost.  
In the following $m_\pi\sim k_F\sim {\cal O}(p)$ are taken of the same chiral order,  and are considered  much smaller than a hadronic scale $\Lambda_\chi$  of several hundreds of MeV that results by integrating out all other particle types, including nucleons with larger three-momentum, heavy mesons and nucleon isobars \cite{wein2}. The final formula obtained in ref.\cite{ref} for the
chiral order $\nu$ of a given diagram is
\be
\nu=4-E+\sum_{i=1}^{V_\pi}(n_i+\ell_i-4)+\sum_{i=1}^V(d_i+v_i+\omega_i-2)+
V_\rho~.
\label{fff}
\ee
where $E$ is the number of external pion lines, $n_i$ is the number of pion lines attached to a vertex without baryons,  $\ell_i$ is the chiral order of the latter with   $V_\pi$ its total number. In addition, $d_i$ is the chiral order of the $i^{th}$
   vertex bilinear in the baryonic fields, $\nu_i$ is the number of mesonic lines attached to it, $\omega_i$ that of only the heavy lines, $V$ is the total number of bilinear vertices and $V_\rho$ is the number of IGVs.
It is important to stress that $\nu$ given in eq.(\ref{fff}) is bounded from below \cite{ref}. Because of the 
last term in eq.(\ref{fff})  adding
 a new IGV to a connected diagram increases the counting at least by 
 one unit.
 The number $ \nu$ given in eq.(\ref{fff}) represents a lower bound for the actual chiral power of a
diagram, $\mu$, so that $\mu\geq \nu$. The real chiral order of a diagram might be different from $\nu$  because 
the nucleon propagators are counted always as ${\cal O}(p^{-2})$ in eq.(\ref{fff}), while for some diagrams there could be propagators  that follow the standard counting. 
 Eq.(\ref{fff}) implies the following conditions for
augmenting the number of lines in a diagram without increasing the chiral power
by adding i) pionic lines attached to mesonic vertices, $\ell_i=n_i=2$, ii) 
 pionic lines attached to meson-baryon vertices, $d_i=v_i=1$ and iii) heavy mesonic lines attached to bilinear vertices, $d_i=0$, $\omega_i=1$.


\section{Nuclear matter energy density}
\label{nmed}

\begin{figure}[t]
\psfrag{Vr=1}{{\tiny $V_\rho=1$}}
\psfrag{Vr=2}{{\tiny $V_\rho=2$}}
\psfrag{Op6}{{\tiny ${\cal O}(p^6)$}}
\psfrag{Op5}{{\tiny $ {\cal O}(p^5)$}}
\psfrag{i}{{\tiny $i$}}
\psfrag{j}{{\tiny $j$}}
\psfrag{q}{{\tiny $q$}}
\psfrag{Leading Order}{\tiny Leading Order}
\psfrag{Next-to-Leading Order}{\tiny Next-to-Leading Order}
\psfrag{eq}{\tiny $\equiv$}
\centerline{\epsfig{file=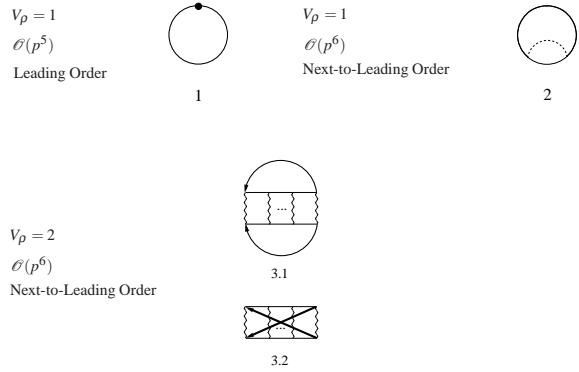,width=.5\textwidth,angle=0}}
\vspace{0.2cm}
\caption[pilf]{\protect \small
Contributions to the nuclear matter energy up 
to  NLO or ${\cal O}(p^6)$. The wiggly lines correspond to the one-pion exchange plus the $NN$ contact interactions from the ${\cal O}(p^0)$ quartic nucleon Lagrangian \cite{wein1,wein2}.
\label{fig:allE}}
\end{figure}

The energy per baryon in nuclear matter at NLO in the counting of Eq.~\eqref{fff} requires to evaluate the set of diagrams shown in Fig.~\ref{fig:allE}. Diagram 1 represents the kinetic energy from the Fermi sea of nucleons. Diagram 2 arises from the nucleon self-energy summed for all the nucleons. Finally, diagrams 3 correspond to the $NN$ interactions in the nuclear medium, 3.1 is for the direct part and 3.2 for the crossed one. 

\begin{figure}[t]
\psfrag{rho}{{\tiny $\begin{array}{l}\\\rho(\rm{fm}^3)\end{array}$}}
\psfrag{Vr=2}{{\small $V_\rho=2$}}
\psfrag{Op6}{{\small ${\cal O}(p^6)$}}
\psfrag{Op5}{{\small $ {\cal O}(p^5)$}}
\psfrag{i}{{\small $i$}}
\psfrag{j}{{\small $j$}}
\psfrag{q}{{\small $q$}}
\psfrag{Leading Order}{Leading Order}
\psfrag{Next-to-Leading Order}{Next-to-Leading Order}
\psfrag{eq}{$\equiv$}
\centerline{\epsfig{file=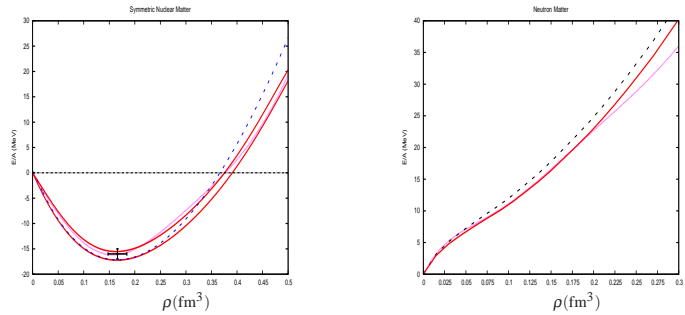,width=.6\textwidth,angle=0}}
\vspace{0.2cm}
\caption[pilf]{\protect \small Energy per baryon for symmetric nuclear matter (left) and pure neutron matter (right).
\label{energy_12}}
\end{figure}

A detailed derivation of the final result is given in Ref.~\cite{anals}. We reproduce here the final expression for the diagrams 3, ${\cal E}_3$, given by
\begin{align}
&{\cal E}_3 =4\sum_{I,J,\ell,S}\sum_{\alpha_1,\alpha_2}(2J+1)\chi(S\ell
I)^2 \int\frac{d^3a}{(2\pi)^3}  \frac{d^3 q}{(2\pi)^3} \theta(\xi_{\alpha_1}-|\val+\vq|)\theta(\xi_{\alpha_2}-|\mathbf{a}-\vq|) \nn\\
& \times 
\Bigl[
-T_{JI}^{i_3}(\ell,\ell,S;\vq^2,\val^2,\vq^2)+ \widetilde{g}_0 \Sigma_{\infty \ell} -
m \int \frac{d^3p}{(2\pi)^3} 
\Bigl\{
\frac{1-\theta(\xi_{\alpha_1}-|\val+\vp|) - \theta(\xi_{\alpha_2}-|\val-\vp|)}{\vp^2-\vq^2-i\ep}
\Sigma_{p\ell}-\frac{\Sigma_{\infty\ell}}{\vp^2}\Bigr\}
\Bigr]~.
\label{e3.reg.2}
\end{align}
where an expansion in $NN$ partial waves in the nuclear medium is used. For further details in the notation we refer to Ref.~\cite{anals}. The only quantity not determined from the $NN$ scattering in vacuum is the constant $\widetilde{g}_0$. However, one can determine its natural size from the way it is introduced \cite{anals}, around $-m m_\pi/4\pi\sim 0.5 10 m_\pi^2$.  The resulting curves for the energy per baryon are shown in Fig.~\ref{energy_12}, left panel for  symmetric nuclear matter and right one for purely neutron matter.  The preferred final values of $\widetilde{g}_0$ are around $-1~m_\pi^2$ for the former and $-0.5~m_\pi^2$ for the latter. We see a good comparison with sophisticated many body calculations \cite{panda} shown by the dotted lines. The experimental point for the symmetric nuclear matter panel is indicated by the crossed. In this case we also are able to reproduce perfectly the nuclear matter incompressibility $K=259$~MeV, to be compared with experiment $K=250\pm  25$~MeV \cite{blaizot}.

\section{Application to neutron stars. Constraining $G_N$.}
\label{gn}

\begin{center}
\begin{figure}[h]
\centerline{\epsfig{file=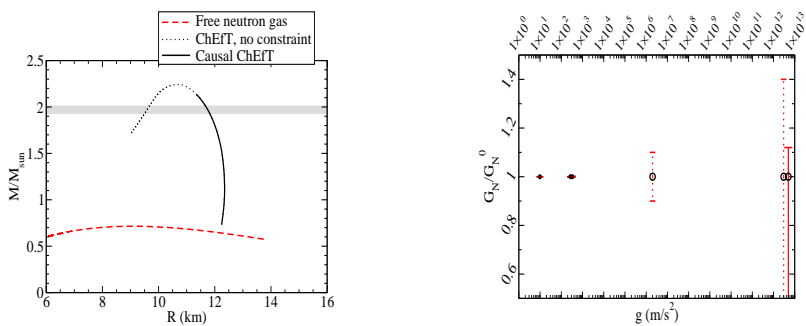,width=.7\textwidth,angle=0}}
\caption{(Color online) Left panel: The neutron star mass as a function of the radius. Right panel: Newtonian constant normalized by its accepted value $6.6738(8)N(m/kg)^2$.  From left to right: laboratory on Earth; orbital determinations of binary pulsars; white dwarf structure; neutron stars with 1.4 solar masses; neutron star with 1.97(4) solar masses.\label{fig:Cavendish}}
\end{figure}
\end{center}

We want to address two question by employing the energy per baryon (neutron matter equation of state at $T=0)$ obtained in the previous section. The first question is to know whether this equation of state, calculated from first principles, is able to account for the large mass of 1.97(4)~$M_\odot$ of the recently observed pulsar J1614-2230 \cite{demoret}. The second question is to know whether the Newtonian constant $G_N$ has the same value as in the Earth for extremely large gravitational fields like those found in the previous pulsar, with its estimated acceleration on the surface around $2\times 10^{12}$~m/s$^{2}$ \cite{felipe}. 
These questions were addressed in Ref.~\cite{felipe} and it was found that our equation of state for neutron matter is able to accommodate such massive neutron stars, with an upper bound for a neutron star mass of around 2.3~$M_\odot$. This is shown in the left panel of Fig.~\ref{fig:Cavendish}.

Regarding $G_N$ the procedure was to use the equation of state of Ref.~\cite{anals} up to Fermi momenta of $450-600$~MeV (leaving this interval as a source of error) and then 
employ the hardest possible equation of state (with a sound velocity equal to the speed of light).  The idea is that if the equation of state is harder this makes $G_N$ to increase in order to provide enough gravitational attraction. Of course, $G_N$ cannot grow indefinitely because otherwise the upper limit value for the mass of the neutron star would decrease too much, being in disagreement with the experimental determination for the pulsar J1614-2230 \cite{demoret}. Taking into account all these considerations for the intense gravitational field in such neutron star, $G_N$ cannot exceed $12\%$ of its value on Earth at $95\%$ confidence level. In the right panel of Fig.~\ref{fig:Cavendish} we show this determination and others from other sources.

\end{document}